\documentclass[11pt]{article}
\usepackage{algorithm}
\usepackage{algorithmic}
\usepackage[algo2e]{algorithm2e}
\RestyleAlgo{ruled}
\usepackage{amsthm,amsmath,amssymb}
\usepackage{natbib}
\usepackage{multirow}
\usepackage[pdftex]{graphicx}
\usepackage{subfigure}
\usepackage{makecell}
\usepackage{booktabs}
\usepackage{array}
\usepackage{fullpage}
\usepackage{url}
\usepackage{bm}
\usepackage{smile}
\usepackage{mathtools}
\usepackage{wrapfig}
\usepackage{lipsum}
\usepackage{mathrsfs}
\usepackage{dsfont}
\usepackage{titling}
\usepackage{epstopdf}
\usepackage{bbm}
\usepackage{relsize}
\usepackage{smile}
\usepackage{natbib}
\usepackage{multirow}
\usepackage{color}
\usepackage{mathtools}
\usepackage{caption}

\usepackage{natbib}
\usepackage{multirow}

\usepackage[usenames,dvipsnames,svgnames,table]{xcolor}
\usepackage[colorlinks=true,
            linkcolor=blue,
            urlcolor=blue,
            citecolor=blue]{hyperref}

\usepackage{comment}
\usepackage{tikz}
\tikzset{
	declare function={
		normcdf(\x,\m,\s)=1/(1 + exp(-0.07056*((\x-\m)/\s)^3 - 1.5976*(\x-\m)/\s));
		zeroone(\x)= (\x<=-2) * (\x*\x + 6*\x + 8)   +
		and(\x>-2, \x<=1) * (2 - \x - \x*\x)     +
		and(\x>1,  \x<=2) * (6 - 8*\x + 2*\x*\x) +
		(\x>2) * (-10 + 6*\x - \x*\x);
	}
}

\usepackage{mathtools}

\usepackage{enumitem}

	\makeatletter
	\newcommand{\mypm}{\mathbin{\mathpalette\@mypm\relax}}
	\newcommand{\@mypm}[2]{\ooalign{%
			\raisebox{.1\height}{$#1+$}\cr
			\smash{\raisebox{-.6\height}{$#1-$}}\cr}}
	\makeatother

	\def\beq{\begin{equation}\begin{aligned}[b]}
	
	\def\eeq{\end{aligned}\end{equation}}

\newtheorem{Lem}{Lemma}
\newtheorem{Th}{Theorem}
\newtheorem{Rem}{Remark}

\def\A{{\bf A}}
\def\a{{\bf a}}
\def\B{{\bf B}}

\def\U{{\bf U}}

\def\X{\bm{X}}
\def\x{\bm{x}}

\def\u{\bm{u}}
\def\0{{\bf 0}}

\def\n{\nonumber}

\def\X{{\bf X}}
\def\x{{\bf x}}
\def\Z{{\bf Z}}
\def\z{{\bf z}}

\def\trans{^{\rm T}}
\def\pr{\hbox{pr}}
\def\wh{\widehat}

\def\var{\hbox{var}}

\def\bse{\begin{eqnarray*}}
\def\ese{\end{eqnarray*}}
\def\be{\begin{eqnarray}}
\def\ee{\end{eqnarray}}
\def\bsq{\begin{equation*}}
\def\esq{\end{equation*}}
\def\bq{\begin{equation}}
\def\eq{\end{equation}}

\def\fyx{f_{Y\mid \X}}

\def\fx{f_{\X}}

\def\fzu{f_{\Z\mid \U}}

\def\fu{f_{\U}}

\def\ora{_{\rm ora}}
\def\para{_{\rm par}}
\def\non{_{\rm non}}

\def\bse{\begin{eqnarray*}}
\def\ese{\end{eqnarray*}}
\def\be{\begin{eqnarray}}
\def\ee{\end{eqnarray}}
\def\bsq{\begin{equation*}}
\def\esq{\end{equation*}}
\def\bq{\begin{equation}}
\def\eq{\end{equation}}

\def\var{\hbox{var}}

\def\wh{\widehat}

\def\eff{_{\rm eff}}

\def\n{\nonumber}

\def\sumI{\sum_{i=1}^N}
\def\sumIP1{\sum_{i=1, i\in P_1}^N}

\def\trans{^{\rm T}}

\def\ba{\boldsymbol\alpha}
\def\bphi{\boldsymbol\phi}

\def\bb{{\boldsymbol\beta}}

\def\0{{\bf 0}}
\def\A{{\bf A}}
\def\U{{\bf U}}

\def\a{{\bf a}}
\def\B{{\bf B}}

\def\f{{\bf f}}
\def\h{{\bf h}}
\def\b{{\bf b}}

\def\m{{\bf m}}

\def\U{{\bf U}}
\def\bS{{\bf S}}

\def\u{{\bf u}}
\def\m{{\bf m}}
\def\v{{\bf v}}

\def\X{{\bf X}}
\def\S{{\bf S}}
\def\s{{\bf s}}
\def\x{{\bf x}}

\def\Z{{\bf Z}}
\def\z{{\bf z}}

\def\bSig{{\bf \Sigma}}

\def\bq{\begin{equation}}
\def\eq{\end{equation}}
\def\pr{\hbox{pr}}
\def\wh{\widehat}

\def\trans{^{\rm T}}

\def\log{{\rm log}}

\def\squarebox#1{\hbox to #1{\hfill\vbox to #1{\vfill}}}

\def\var{\hbox{var}}

\def\bse{\begin{eqnarray*}}
\def\ese{\end{eqnarray*}}
\def\be{\begin{eqnarray}}
\def\ee{\end{eqnarray}}
\def\bsq{\begin{equation*}}
\def\esq{\end{equation*}}
\def\bq{\begin{equation}}
\def\eq{\end{equation}}
\def\wh{\widehat}

\def\trans{^{\rm T}}

\def\logit{{\mbox{logit}}}
\def\boxit#1{\vbox{\hrule\hbox{\vrule\kern6pt\vbox{\kern6pt#1\kern6pt}\kern6pt\vrule}\hrule}}

\def\trans{^{\rm T}}
\def\pr{\hbox{pr}}
\def\wh{\widehat}

\def\sumI{\sum_{i=1}^N}

\def\fyx{f_{Y\mid \X}}

\def\fx{f_{\X}}

\def\fzu{f_{\Z\mid \U}}

\def\fu{f_{\U}}

\numberwithin{equation}{section}

\begin{document}

\title{A Versatile Estimation Procedure without Estimating the
  Nonignorable Missingness Mechanism}

\author{Jiwei Zhao\thanks{Department of Biostatistics, State University of New York at Buffalo, Buffalo, NY 14214, USA; e-mail: \texttt{zhaoj@buffalo.edu}.}~~~~~Yanyuan Ma\thanks{Department of Statistics, Pennsylvania State University, University Park, PA 16802, USA; e-mail: \texttt{yzm@psu.edu}.}
}
\date{\today}

\maketitle

\vspace{-0.3in}

\begin{abstract}
We consider the estimation problem in a regression setting where the
outcome variable is subject to nonignorable missingness and
identifiability is ensured by the shadow variable approach.
We propose a versatile estimation procedure where
modeling of missingness mechanism is completely bypassed.
We show that our estimator is easy to implement
and we derive the asymptotic theory of the proposed estimator.
We also investigate some alternative estimators under different scenarios.
Comprehensive simulation studies are conducted to
demonstrate the finite sample performance of the method.
We apply the estimator to a children's mental health study to illustrate its usefulness.
\end{abstract}

\section{Introduction}

In statistical data analysis, the issue of missing values
is a rule rather than an exception. There are often many
missing data in biomedical and health related studies, social sciences
and survey sampling. How to appropriately address missingness
is fascinating but challenging, and has drawn much
attention to statisticians in the past
several decades.

In the missing data literature, the missingness mechanism is a key
concept and a fundamental and useful taxonomy to
distinguish different problems. The missingness is named ignorable if
it depends on the observed data only; otherwise, it is named
nonignorable. Rich literatures exist on handling
ignorable missing data \citep{rubin1987multiple, robins1994estimation,
  schafer1997analysis, little2002statistical,
  tsiatis2006semiparametric, kim2013statistical,
  molenberghs2014handbook}.
However, in many practical situations, it is highly likely that the
missingness actually also depends on the missed variables themselves hence is nonignorable.
Research for nonignorable missing data
is not yet as complete due to its difficulties.
Simply applying existing methods for ignorable missing data to nonignorable ones
may lead to biased parameter estimation, incorrect
standard errors and, as a consequence, incorrect statistical inference and conclusions.

One notorious issue for analyzing nonignorable missing data is the
model identifiability. Here, identifiability means that any two
different sets of parameters produce two different models. In the
literature, different strategies \citep{tchetgen2017general,
sun2017semiparametric} are used to achieve identifiability. Here
we adopt the shadow variable strategy, popularly used and well
documented in \cite{wang2014instrumental}, \cite{zhao2015semiparametric} and
\cite{miao2016varieties}. More details of the shadow variable strategy and
its treatment in applications are presented in Section
\ref{sec:prep}.

The most controversial part in analyzing nonignorable missing data is
on modeling the missingness mechanism. Because of its dependence on
the unobserved data, it is nearly impossible to verify the mechanism
model in practice except for a few special scenarios \citep{d2010new}. In
the literature, there are many parametric modeling attempts for the
mechanism\citep{ibrahim1996parameter,
  rotnitzky1997analysis, qin2002estimation, chang2008using,
  wang2014instrumental, morikawa2016semiparametric}, but parametric
mechanism model is generally considered to be restrictive.
 \cite{kim2011semiparametric} and \cite{shao2016semiparametric}
 extended the parametric mechanism to a semiparametric framework which
 contains a more flexible nonparametric component. However
   these  semiparametric
 mechanism models are also confined to a special structure and can still be misspecified.

Due to the difficulty in modeling the missingness mechanism,
in this paper we completely avoid this practice. We propose a
versatile estimation procedure which does not require modeling or
estimating the missingness mechanism.
The key idea of our proposal is to view the mechanism as a nuisance
parameter in a semiparametric model, and to project away its effect via
semiparametric treatment \citep{bickel1993efficient,
  tsiatis2006semiparametric}. In the estimator we construct in this work,
only a working model for the mechanism is needed in the
implementation, and the working model does not have to contain the true
mechanism.

Our procedure requires estimating integrals depending on the
probability density function (pdf) or probability mass function (pmf)
of the covariate variable. Because covariates are fully observed,
this is a complete-data problem and many statistical methods exist in
the literatures. We propose to estimate the integral through empirical
expectation if the integral can be viewed as a marginal expectation,
and to estimate it through nonparametric regression technique, such
as kernels, if the integral can be converted into a conditional
expectation. Our procedure is more robust compared to parametric
estimation of the pdf/pmf, and is
simpler to implement compared to nonparametric estimation.
It is also worthwhile to mention that it is technically challenging to establish
the asymptotic theory of the proposed estimator, which requires
extensive use of bilinear operators in combination with semiparametric
treatments.

The rest of the paper is as follows. In Section \ref{sec:prep}, we
clarify notations and assumptions, describe the shadow variable
strategy and lay down the model identification conditions. In Section
\ref{sec:special}, we study the situation where the whole
covariate vector serves as the shadow variable. We derive
the efficient score,  propose our estimator and establish the
asymptotic theory. The
parallel results under the more general situation where part
of the covariate serves as the shadow variable is established in Section
\ref{sec:general}.
A few alternative estimators under difference scenarios are investigated in Section \ref{sec:other}.
In Section \ref{sec:simu}, we conduct comprehensive
simulation studies to demonstrate the finite sample performance of our
proposed methods under various situations. In Section \ref{sec:data},
we analyze a data concerning a children's mental health study. The
paper is concluded with a discussion in Section \ref{sec:disc}.

\section{Notations and Assumptions}\label{sec:prep}

Consider the regression model $\fyx(y,\x)=\fyx(y,\x;\bb)$,
where $\bb$ is a
$p$-dimensional unknown parameter to be estimated. The covariate $\X$
is fully observed and let the pdf/pmf of $\X$ be $\fx(\cdot)$.
The response variable $Y$ is subject to missingness. Let the binary
variable $R$ be the missingness indicator, with $R=1$ for an observed
$Y$ and $R=0$ for a missing $Y$. Write the missingness mechanism as
$\pr(R=1\mid Y,\X)$.
We observe $N$ independent and identically distributed
realizations of $(R,RY,\X)$, written as
$\{(r_i, r_iy_i, \x_i)\}, i=1,\ldots,N$. Without loss of generality,
we assume that the first $n$ subjects are completely observed, i.e. $r_i=1$
for $i=1, \dots, n$, while the
remaining $N-n$ subjects have $r_i=0$ for $i=n+1, \dots, N$.

We adopt the shadow variable framework, i.e., we assume that the
covariate $\X$ can be decomposed as $\X=(\U\trans,\Z\trans)\trans$ and
\be
\pr(R=1\mid y,\x)=\pr(R=1\mid y,\u)=\pi(y,\u).
\label{eq:general}
\ee
The variable $\Z$ is termed the shadow variable.
This implies that part of the covariate, $\Z$, is independent of the
missingness indicator $R$ conditional on the response $Y$ and the
other part of the covariate $\U$. Consequently,
while $\Z$ appears in the
model $\fyx(y,\x)$, it does not in the model $\pr(R=1\mid Y,\X)$,
hence is shadowed out. Note that $\Z$ can be $\X$, hence the whole
covariate $\X$ itself is the shadow variable, but $\Z$ cannot be empty,
which degenerates to the no shadow variable situation.
The shadow variable assumption is popularly used in the literature
\citep{wang2014instrumental, zhao2015semiparametric, miao2016varieties} and is found to be
very useful in applications. For example, in analyzing the nonignorable
missing outcome data from the children's mental health study
\citep{ibrahim2001using}, the authors implemented a parametric EM
algorithm and found that the mechanism model does not depend on a
binary covariate variable representing whether a father figure is present in the household or not, hence the authors removed this
covariate in their missingness mechanism model in their subsequent
analysis. Here, the covariate regarding the presence of a father
figure is a shadow variable.
Given the shadow variable assumption, we find that all unknown
components become identifiable, as we state in Lemma \ref{lem:iden},
with its proof  in Appendix. 

\begin{Lem}[Identifiability]\label{lem:iden}
Under the shadow variable assumption (\ref{eq:general}), $\bb$,
  $\pi(y,\u)$ and $\fx(\u,\z)$ are  identifiable.
\end{Lem}

If we adopt a likelihood approach, even though our sole interest is in
estimating $\bb$, we cannot avoid the estimation of both $\pi(y,\u)$
and $\fx(\u,\z)$. While the estimation of $\fx(\u,\z)$ is a standard
problem  since there is no missing data in the variable $\X$,
the estimation of $\pi(y,\u)$ is challenging.
Due to the missingness in $Y$, the $\pi(\cdot)$ model is usually unverifiable and
can be easily misspecified in practice.

Aware of this difficulty, we propose to estimate $\bb$
while avoiding modeling or estimating the missingness mechanism.
Instead, we only need to posit a working model for $\pi(y,\u)$, which
could be  misspecified.
We show that, using an arbitrary working model
$\pi^\ast(y,\u)$, our estimator of $\bb$ is always consistent and
asymptotically normal, hence our procedure is robust to mechanism
misspecification.

For ease of illustration, also with its own importance, in Secton
\ref{sec:special}
we first analyze a special case of
(\ref{eq:general}) where the whole covariate serves as the
shadow variable, i.e. $\X=\Z$ and
\be
\pr(R=1\mid y,\x) = \pr(R=1\mid y)=\pi(y).
\label{eq:special}
\ee
Analysis under the general assumption
(\ref{eq:general}), which turns out to be statistically very
different and
mathematically more challenging, is
conducted in Section \ref{sec:general}.

\section{Proposed Estimator under Special Assumption
  (\ref{eq:special})}\label{sec:special}

\subsection{Estimation Procedure}\label{sec:est1}

Under (\ref{eq:special}), the joint pdf of $(\X, RY, R)$ is
\bse
\fx(\x) \{\fyx(y,\x;\bb)\pi(y)\}^{r}
\left\{1-\int \fyx(t,\x;\bb)\pi(t)dt\right\}^{1-r}.
\ese
Because $\bb$ is the parameter of interest while $\fx(\x)$ and $\pi(y)$
are nuisance, we take a semiparametric approach and derive the
nuisance tangent space, its orthogonal complement and the
efficient score with respect to $\bb$. In Appendix,
we first derive that
the nuisance tangent space $\Lambda = \Lambda_{\fx}\oplus \Lambda_{\pi}$, where
\bse
\Lambda_{\fx}&=&[\a(\x): E\{\a(\X)\}=\0],\\
\Lambda_\pi&=& [
r\b(y)-(1-r)\frac{\int \fyx(t,\x;\bb)\b(t)\pi(t)dt}{\int
  \fyx(t,\x;\bb)\{1-\pi(t)\}dt}:\forall \b(y)],
\ese
where $\oplus$ stands for the addition of two spaces
that are orthogonal to each other. We also derive
the orthogonal complement of $\Lambda$ to be
\bse
\Lambda^\perp&=&\left[\a(\x,r,ry):
E\{\a(\x,R,RY)\mid\x\}=\0, \forall \b(Y),\right.\\ \nonumber
&&\left.E\left\{
\a(\X,R,RY)R\b(Y)-\frac{\a(\X,R,RY) (1-R)\int \fyx(t,\x;\bb)\b(t)\pi(t)dt}{1-\int
  \fyx(t,\x;\bb)\pi(t)dt}\right\}=\0
\right].
\ese

The form of $\Lambda^\perp$ permits many possibilities for
constructing estimating equations for $\bb$. Among all elements in
$\Lambda^\perp$, the most interesting one is the efficient
score, defined as the orthogonal projection of the score vector
$\S_\bb$ onto $\Lambda^\perp$, where
\be\label{eq:score}
\S_\bb(\x,r,ry,\bb)
=r\frac{\f_\bb(y,\x;\bb)}{\fyx(y,\x;\bb)}
-(1-r)\frac{\int \f_\bb(t,\x;\bb)\pi(t)dt}{\int
  \fyx(t,\x;\bb)\{1-\pi(t)\}dt}.
\ee
Here $\f_\bb(y,\x;\bb)\equiv\partial \fyx(y,\x;\bb)/\partial\bb$.
In Appendix, 
we show that
\bse
&&\S\eff(\x,r,ry)\\
  &=& \frac{r \f_\bb(y,\x;\bb)}{\fyx(y,\x;\bb)}
-\frac{(1-r)\int \f_\bb(t,\x;\bb)\pi(t)dt}{\int
  \fyx(t,\x;\bb)\{1-\pi(t)\}dt}
  -r\b(y)
+\frac{(1-r)\int\b(t)\fyx(t,\x;\bb)\pi(t)dt}{\int
  \fyx(t,\x;\bb)\{1-\pi(t)\}dt},
\ese
where
$\b(y)$ is the solution to the integral equation
\be\label{eq:intspecial}
&&\int \left\{\f_\bb(y,\x;\bb)+\frac{\int\f_\bb(t,\x;\bb)\pi(t)dt}{1-\int
  \fyx(t,\x;\bb)\pi(t)dt}\fyx(y,\x;\bb)\right\}\fx(\x) d\mu(\x)\\
&=&
\int\left\{\b(y)\fyx(y,\x;\bb)+\frac{\int\b(t)\fyx(t,\x;\bb)\pi(t)dt}{1-\int
  \fyx(t,\x;\bb)\pi(t)dt}\fyx(y,\x;\bb)\right\}\fx(\x) d\mu(\x).\n
\ee

Despite of the results above,
the efficient score $\S\eff$ is not readily implementable because it
contains the unknown quantities $\fx(\x)$ and $\pi(y)$.
As we have pointed out, we aim to avoid estimating or even modeling
$\pi(y)$. Thus, we propose to adopt a working model of the mechanism,
denoted $\pi^\ast(y)$. We show in Appendix 
that in the construction of $\bS\eff(\x,r,ry)$, we can adopt $\pi^\ast(y)$ and
the resulting ``working model based efficient score'' $\bS\eff^\ast(\x,r,ry)$ still has
mean zero.
On the other hand, the integrations in (\ref{eq:intspecial}) can be
viewed as expectations with respect to the covariate $\X$.
Because  $\X$ is
fully observed, we recommend to approximate the expectations using
their corresponding empirical versions.
Combining these two aspects, we propose the following
flexible estimation procedure.

{\noindent \underline{\bf Algorithm 1: Algorithm under Special Assumption (\ref{eq:special})}
\begin{enumerate}
 \item[Step 1.]
Posit a working model for $\pi(y)$, denote it $\pi^\ast(y)$.

  \item[Step 2.]
Obtain
$\wh\b^\ast(y,\bb)$ by solving the integral equation
\be\label{eq:appintspecial}
&&\frac1N\sumI \left\{\f_\bb(y,\x_i;\bb)+\frac{\int\f_\bb(t,\x_i;\bb)\pi^\ast(t)dt}{1-\int
  \fyx(t,\x_i;\bb)\pi^\ast(t)dt}\fyx(y,\x_i;\bb)\right\}\\
&=&\frac1N
\sumI\left\{\b(y)\fyx(y,\x_i;\bb)+\frac{\int\b(t)\fyx(t,\x_i;\bb)\pi^\ast(t)dt}{1-\int
  \fyx(t,\x_i;\bb)\pi^\ast(t)dt}\fyx(y,\x_i,\bb)\right\}.\n
\ee

  \item[Step 3.] Insert $\wh\b^\ast(y,\bb)$ into the efficient score
    expression to obtain
  \bse
  \bS\eff^\ast\{\x,r,ry,\bb,\wh\b^\ast(\cdot,\bb)\}
&=&
\frac{r \f_\bb(y,\x;\bb)}{\fyx(y,\x;\bb)}
-\frac{(1-r)\int \f_\bb(t,\x;\bb)\pi^\ast(t)dt}{\int
  \fyx(t,\x;\bb)\{1-\pi^\ast(t)\}dt}
 -r\wh\b^\ast(y,\bb)\\
&&+\frac{(1-r)\int\wh\b^\ast(t,\bb)\fyx(t,\x;\bb)\pi^\ast(t)dt}{\int
  \fyx(t,\x;\bb)\{1-\pi^\ast(t)\}dt}.
\ese

\item[Step 4.]
Solve the estimating equation $\sumI \bS\eff^\ast\{\x_i,r_i,r_i
y_i,\bb,\wh\b^\ast(\cdot,\bb)\}=\0$ to obtain
the estimator $\wh\bb$.
\end{enumerate}

In Step 2 of Algorithm 1,  (\ref{eq:appintspecial})
is a Fredholm integral equation of the
second type and is well-posed, hence we obtain $\wh\b^\ast(y,\bb)$
using the method proposed in
\cite{atkinson1976automatic}.

\subsection{Theoretical Property}\label{sec:asym1}

To theoretically analyze $\wh\bb$,
the technical difficulties mainly stem from quantifying the difference
between the solutions of the integral equations
(\ref{eq:intspecial}) and (\ref{eq:appintspecial}).
To proceed, we first introduce some notation.
We define
\bse
u_1(y)&=&E\left\{\fyx(y,\X_i;\bb)\right\}
=\int \fyx(y,\x;\bb) f_\X(\x)d\mu(\x), \\
u_2(t,y)&=&E\left\{\frac{ \fyx(y,\X_i;\bb) \fyx(t,\X_i;\bb)}{1-\int
  \fyx(t,\X_i;\bb)\pi^\ast(t)dt}\right\}\pi^\ast(t),\\
\v(y)&=&E \left\{\f_\bb(y,\X_i;\bb)+\frac{\int\f_\bb(t,\X_i;\bb)\pi^\ast(t)dt}{1-\int
  \fyx(t,\X_i;\bb)\pi^\ast(t)dt}\fyx(y,\X_i;\bb)\right\},
\ese
and the linear operation ${\cal A}(\cdot, y)$
 on $\b(\cdot)$ as
\bse
{\cal A}(\b)(y)
\equiv\b(y)u_1(y)+\int\b(t)u_2(t,y)dt.
\ese
Similarly, let
\bse
u_{1i}(y)&=&
\fyx(y,\x_i;\bb),\\
u_{2i}(t,y)&=&
\frac{\fyx(y,\x_i;\bb)\fyx(t,\x_i;\bb)}{1-\int
  \fyx(t,\x_i;\bb) \pi^\ast(t)dt}\pi^\ast(t),\\
\v_i(y)&=&\f_\bb(y,\x_i;\bb)+\frac{\int\f_\bb(t,\x_i;\bb)\pi^\ast(t)dt}{1-\int
  \fyx(t,\x_i;\bb)\pi^\ast(t)dt}\fyx(y,\x_i;\bb).
\ese
Note that $u_{1i},u_{2i},\v_i$ depend on the $i$th observation only
through $\x_i$.
Also define
\bse
\wh u_1(y)=N^{-1}\sumI u_{1i}(y),\ \ \ \
\wh u_2(t,y)=N^{-1}\sumI u_{2i}(t,y),\ \ \ \
\wh\v(y)=N^{-1}\sumI \v_i(y).
\ese
Similar to ${\cal A}$, we define the linear operator
\bse
{\cal A}_i(\b)(y)\equiv
\b(y)u_{1i}(y)+\int \b(t) u_{2i}(t,y)dt, \ \ \ \
\wh {\cal A}(\b)(y)
\equiv N^{-1}\sumI{\cal A}_i(\b)(y).
\ese
We also introduce some regularity conditions.
\begin{itemize}
  \item[(A1)] $0<\delta<\pi^\ast(t)<1-
\delta$ for all $t$, where $0<\delta<1/2$ is a constant.
  \item[(A2)] The true parameter value of $\bb$ belongs to a bounded domain.
The support sets of $\fx(\x), f_Y(y), \pi(y)$ are compact.
  \item[(A3)]  The functions $u_1(y), u_2(t,y)$
are bounded and bounded away from zero on their support. The
score function $\S_\bb(\x,y;\bb)\equiv
\f_\bb(y,\x;\bb)/\fyx(y,\x;\bb)$ is bounded, hence its orthogonal
projection $\b^\ast(y)$ is also
bounded.
\end{itemize}

Under these regularity conditions, the following result, with its
proof in Appendix,
guarantees that $\|{\cal
  A}(\b)\|_\infty$ is well bounded by $\|\b\|_\infty$.
\begin{Lem}\label{lem:calA}
Under the regularity conditions (A1)-(A3), there exist constants $0<c_1<c_2<\infty$ such that
$c_1\|\b\|_\infty\le\|{\cal A}(\b)\|_\infty\le
c_2\|\b\|_\infty$.
\end{Lem}

Further, we have the following result, with its proof given in Appendix, 
concerning the asymptotic distribution of $\wh\bb$.
\begin{Th}\label{th:emp}
  For any choice of $\pi^\ast(y)$, under Conditions (A1)-(A3), $\wh\bb$ satisfies
  \bse
  \sqrt{N}(\wh\bb - \bb) \to N\{\0, \A^{-1}\B(\A^{-1})\trans\},
  \ese
  in distribution when $N\to\infty$,
  where
  \bse
  \A&=&E\left[\frac{d\bS\eff^\ast\{\X_i,R_i,R_iY_i,\bb,\b^\ast(\cdot,\bb)\}}{d\bb\trans}\right],\\
\B&=&\var[\bS\eff^\ast\{\X_i,R_i,R_iY_i,\bb,\b^\ast(\cdot,\bb)\}-\h(\X_i)],\\
\h(\x_i)&=&
\int\left[\{\pi(y)-1\}\{
\v_i(y)-{\cal A}_i(\b^\ast)(y)\}+{\cal A}^{-1}\{\v_i- {\cal A}_i(\b^\ast)\}(y)u_1(y)
\right]d\mu(y).
\ese
\end{Th}
Here
\bse
&&\frac{d\bS\eff^\ast\{\X_i,R_i,R_iY_i,\bb,\b^\ast(\cdot,\bb)\}}{d\bb\trans}\\
&\equiv&
\frac{\partial\bS\eff^\ast\{\X_i,R_i,R_iY_i,\bb,\b^\ast(\cdot,\bb)\}}{\partial\bb\trans}
+\frac{\partial\bS\eff^\ast\{\X_i,R_i,R_iY_i,\bb,\b^\ast(\cdot,\bb)\}}{\partial{\b^\ast}\trans}\frac{\partial
  \b^\ast(\cdot,\bb)}{\partial\bb\trans}.
\ese

\begin{Rem}\label{rem:emp}
One can easily verify that $E\{\v_i-{\cal A}_i (\b^\ast)\} (y)=\0$ and
$E{\cal A}^{-1}\{\v_i-{\cal A}_i (\b^\ast)\} (y)=\0$, hence
$\h(\x_i)\in \Lambda_{\fx}$. Thus, if fortunately the working model
$\pi^*(y)$ is chosen as the true mechanism $\pi(y)$, then $E(\bS\eff
\h\trans) =
  0$ because $\bS\eff \in \Lambda^\perp$. This means $\B$ in Theorem
  \ref{th:emp} can be further simplified to $\B =
  \var[\bS\eff\{\X_i,R_i,R_iY_i,\bb,\b(\cdot,\bb)\}]+\var\{\h(\X_i)\}$ under this situation. Thus,
  we can view $h(\x_i)$ as the additional term to account for the cost from
  empirical approximation of the integrals in (\ref{eq:intspecial}).
\end{Rem}

\section{Proposed Estimator under General Assumption (\ref{eq:general})}
\label{sec:general}


Under the general model  (\ref{eq:general}), the joint pdf of
$(\X, RY, R)$ is
\bse
\fx(\u,\z) \{\fyx(y,\u,\z;\bb)\pi(y,\u)\}^{r}
\left\{1-\int \fyx(t,\u,\z;\bb)\pi(t,\u)dt\right\}^{1-r},
\ese
where $\bb$ is still the parameter of interest, and the functions
$\fx(\cdot)$ and $\pi(\cdot)$ are the nuisance parameters.
Because
the mechanism model $\pi(y,\u)$ now also
depends on partial covariate $\u$, the situation is much different
from that considered in Section \ref{sec:special}. We in fact
show in Appendix 
that the nuisance tangent space orthogonal complement in this case is
\bse
\Lambda^\perp=\left\{\a(\u,\z,r,ry):
E\{\a(\u,\z,R,RY)\mid\u,\z\}=\0,
E\left[
\{\a(\u,\Z,1,y) -\a(\u,\Z,0,0)\} \mid y,\u\right]
=\0
\right\},
\ese
and the efficient score for parameter $\bb$ is
\bse
\bS\eff(\u,\z,r,ry)
&=&\frac{r \f_\bb(y,\u,\z;\bb)}{\fyx(y,\u,\z;\bb)}
-\frac{(1-r)\int \f_\bb(t,\u,\z;\bb)\pi(t,\u)dt}{\int
  \fyx(t,\u,\z;\bb)\{1-\pi(t,\u)\}dt}\\
&&  -r\b(y,\u)
+\frac{(1-r)\int\b(t,\u)\fyx(t,\u,\z;\bb)\pi(t,\u)dt}{\int
  \fyx(t,\u,\z;\bb)\{1-\pi(t,\u)\}dt},
\ese
where $\f_\bb(y,\u,\z;\bb)\equiv\partial
\fyx(y,\u,\z;\bb)/\partial\bb$,
and $\b(y,\u)$ satisfies
the integral equation
\be\label{eq:intgeneral}
&&\int
\left[
\frac{\f_\bb(y,\u,\z;\bb)}{\fyx(y,\u,\z;\bb)}
 -\b(y,\u)
+\frac{\int \f_\bb(t,\u,\z;\bb)\pi(t,\u)dt}{\int
  \fyx(t,\u,\z;\bb)\{1-\pi(t,\u)\}dt}\right.\n\\
&&\left.  -\frac{\int\b(t,\u)\fyx(t,\u,\z;\bb)\pi(t,\u)dt}{\int
  \fyx(t,\u,\z;\bb)\{1-\pi(t,\u)\}dt}\right]
\fzu(\z,\u)
\fyx(y,\u,\z;\bb)d\mu(\z)
=\0.
\ee
Note that we used the decomposition $\fx(\u,\z)=\fzu(\z,\u)\fu(\u)$ above.
We can see that the space $\Lambda^\perp$ has quite different form from
its counterpart in Section \ref{sec:special}, caused by
the additional inclusion of $\U$ in the mechanism model. Nevertheless,
in Appendix 
we verify that a misspecified
$\pi(y,\u)$ model, $\pi^*(y,\u)$, can be
employed in the construction of $\bS\eff(\u,\z,r,ry)$ and the mean zero property of the efficient score will still
be retained.

In an effort to construct an estimator similar in spirit to
$\wh\bb$ in Section \ref{sec:special}, we realize that
we would have to handle the $\U$ part and the $\Z$ part of the covariates
differently because they play
different roles. In fact, while we could be totally ``empirical'' with
respect to $\Z$, we would have to remain ``nonparametric'' with respect to
$\U$.
Specifically,
recognizing that the left hand side of (\ref{eq:intgeneral}) is a
conditional expectation, we
approximate the integral equation (\ref{eq:intgeneral}) by
\be
&&\frac{1}{N}\sumI\left[
\frac{\f_\bb(y,\u,\z_i;\bb)}{\fyx(y,\u,\z_i;\bb)} -\b(y,\u)
+\frac{\int \f_\bb(t,\u,\z_i;\bb)\pi^\ast(t,\u)dt}{\int
  \fyx(t,\u,\z_i;\bb)\{1-\pi^\ast(t,\u)\}dt}\right.\n\\
&&\left.  -\frac{\int\b(t,\u)\fyx(t,\u,\z_i;\bb)\pi^\ast(t,\u)dt}{\int
  \fyx(t,\u,\z_i;\bb)\{1-\pi^\ast(t,\u)\}dt}\right]
\fyx(y,\u,\z_i;\bb)K_h(\u_i-\u)=\0,\label{eq:appintgeneral}
\ee
utilizing the nonparametric regression technique, where $K_h(\cdot)=K(\cdot/h)/h$, $K(\cdot)$ is a
kernel function and $h$ is a bandwidth, with their conditions detailed later.
Once $\wh\b^\ast(t,\u)$ is obtained from solving
 (\ref{eq:appintgeneral}), we can then
proceed to construct the estimating equation and obtain the
estimator. For completeness, we write out the algorithm.

{\noindent \underline{\bf Algorithm 2: Algorithm under General Assumption (\ref{eq:general})}}
\begin{enumerate}

  \item[Step 1.]
Posit a working model for $\pi(y,\u)$, denote it $\pi^\ast(y,\u)$.

\item[Step 2.]
Obtain
$\wh\b^\ast(y,\u,\bb)$ by solving the  integral equation
(\ref{eq:appintgeneral}).

  \item[Step 3.] Insert $\wh\b^\ast(y,\u,\bb)$ into the  efficient score expression to obtain
  \bse
\bS\eff^\ast\{\u,\z,r,ry,\bb,\wh\b^*(\cdot,\u,\bb)\}&=&\frac{r \f_\bb(y,\u,\z;\bb)}{\fyx(y,\u,\z;\bb)}
-\frac{(1-r)\int \f_\bb(t,\u,\z;\bb)\pi^\ast(t,\u)dt}{\int
  \fyx(t,\u,\z;\bb)\{1-\pi^\ast(t,\u)\}dt}\\
&&  -r\wh\b^\ast(y,\u,\bb)
+\frac{(1-r)\int\wh\b^\ast(t,\u,\bb)\fyx(t,\u,\z;\bb)\pi^\ast(t,\u)dt}{\int
  \fyx(t,\u,\z;\bb)\{1-\pi^\ast(t,\u)\}dt}.
\ese

\item[Step 4.]
Solve the estimating equation $\sumI \bS\eff^\ast\{\u_i,\z_i,r_i,r_i
y_i,\bb,\wh\b^*(\cdot,\u_i,\bb)\}=\0$ to obtain
the estimator for $\bb$. We still denote as $\wh\bb$.
\end{enumerate}

Like (\ref{eq:appintspecial}),
(\ref{eq:appintgeneral})
is also a Fredholm integral equation of the second type and is
well-posed, so
it can also be solved by the method proposed in
\cite{atkinson1976automatic}.
Note that only $\wh \b^*(y,\u_i,\bb)$'s are needed in Algorithm 2, instead
of the generic function $\wh \b^*(y,\u,\bb)$.

To study the theoretical property of $\wh\bb$,
the technical difficulties mainly stem from quantifying the difference
between the solutions of the integral equations
(\ref{eq:intgeneral}) and (\ref{eq:appintgeneral}).
To proceed, we first introduce some notation.
We define
\bse
u_1(y,\u)&=&\fu(\u)E\left\{\fyx(y,\u,\Z_i;\bb)\mid\U=\u\right\}
=\fu(\u)\int \fyx(y,\u,\z;\bb) \fzu(\z,\u)d\mu(\z), \\
u_2(t,y,\u)&=&\fu(\u)E\left\{\frac{ \fyx(y,\u,\Z_i;\bb) \fyx(t,\u,\Z_i;\bb)}{1-\int
  \fyx(t,\u,\Z_i;\bb)\pi^\ast(t,\u)dt}\mid \U=\u\right\}\pi^\ast(t,\u),\\
\v(y,\u)&=&\fu(\u)E \left\{\f_\bb(y,\u,\Z_i;\bb)+\frac{\int\f_\bb(t,\u,\Z_i;\bb)\pi^\ast(t,\u)dt}{1-\int
  \fyx(t,\u,\Z_i;\bb)\pi^\ast(t,\u)dt}\fyx(y,\u,\Z_i;\bb)\mid\U=\u\right\},
\ese
and the linear operation ${\cal A}(\cdot, y,\u)$
 on $\b(\cdot)$ as
\bse
{\cal A}(\b)(y,\u)
\equiv\b(y,\u)u_1(y,\u)+\int\b(t,\u)u_2(t,y,\u)dt.
\ese
Similarly, let
\bse
u_{1i}(y,\u)&=&K_h(\u_i-\u)
\fyx(y,\u,\z_i;\bb),\\
u_{2i}(t,y,\u)&=&K_h(\u_i-\u)
\frac{\fyx(y,\u,\z_i;\bb)\fyx(t,\u,\z_i;\bb)}{1-\int
  \fyx(t,\u,\z_i;\bb) \pi^\ast(t,\u)dt}\pi^\ast(t,\u),\\
\v_i(y,\u)&=&K_h(\u_i-\u) \left\{\f_\bb(y,\u,\z_i;\bb)+\frac{\int\f_\bb(t,\u,\z_i;\bb)\pi^\ast(t,\u)dt}{1-\int
  \fyx(t,\u,\z_i;\bb)\pi^\ast(t,\u)dt}\fyx(y,\u,\z_i;\bb)\right\}.
\ese
Note that $u_{1i},u_{2i},\v_i$ depend on the $i$th observation only
through $\x_i$.
Also define
\bse
\wh u_1(y,\u)=N^{-1}\sumI u_{1i}(y,\u),\ \ \ \
\wh u_2(t,y,\u)=N^{-1}\sumI u_{2i}(t,y,\u),\ \ \ \
\wh\v(y,\u)=N^{-1}\sumI \v_i(y,\u).
\ese
Similar to ${\cal A}$, we define the linear operator
\bse
{\cal A}_i(\b)(y,\u)\equiv
\b(y,\u)u_{1i}(y,\u)+\int \b(t,\u) u_{2i}(t,y,\u)dt, \ \ \ \
\wh {\cal A}(\b)(y,\u)
\equiv N^{-1}\sumI{\cal A}_i(\b)(y,\u).
\ese

We need the following
conditions on the true functions, kernel function and the bandwidth.
\begin{itemize}
  \item[(B1)] $0<\delta<\pi^\ast(t,\u)<1-
\delta$ for all $(t,\u)$, where $0<\delta<1/2$ is a constant.
  \item[(B2)] The true parameter value of $\bb$ belongs to a bounded domain.
The support sets of $\fzu(\z,\u), f_Y(y), \pi(y,\u)$ are compact.
  \item[(B3)]  The functions $u_1(y,\u), u_2(t,y,\u)$
are bounded and bounded away from zero on their support. The
score function $\S_\bb(\u,\z,y;\bb)\equiv
\f_\bb(y,\u,\z;\bb)/\fyx(y,\u,\z;\bb)$ is bounded, hence its orthogonal
projection $\b^\ast(y,\u)$ is also
bounded.

\item[(B4)] The univariate kernel
function $K(\cdot)$ is bounded and symmetric,
has a bounded derivative and compact support $[-1, 1]$, and
satisfies $\int K(u)du=1$,
$\mu_m = \int u^m K(u)du \neq 0$,
$\int u^r K(u)du= 0$ for $r=1, \dots, m-1$.
 $K_h(u) = K(u/h)/h$. The
$d$-dimensional kernel function is a product of $d$ univariate kernel
functions, i.e., $K(\u)=\prod_{j=1}^d K(u_j)$, and
$K_h(\u)=\prod_{j=1}^d K_h(u_j)=h^{-d}\prod_{j=1}^d K(u_j/h)$ for
$\u=(u_1, \ldots, u_d)\trans$ and bandwidth $h$. Here $d$ is the
dimension of $\u$.

\item[(B5)] The bandwidth $h$ satisfies $h\to0$, $Nh^{2d}\to\infty$
  and $Nh^{2m}\to0$.

\end{itemize}

Under these regularity conditions, we have the following Lemma
to ensure that $\|{\cal
  A}(\b)\|_\infty$ is  bounded by $\|\b\|_\infty$. Its proof is
in Appendix, 
\begin{Lem}\label{lem:calAnonyu}
Under the regularity conditions (B1)-(B3), there exist constants $0<c_1<c_2<\infty$ such that
$c_1\|\b\|_\infty\le\|{\cal A}(\b)\|_\infty\le
c_2\|\b\|_\infty$.
\end{Lem}

The theoretical property of $\wh\bb$ is summarized in Theorem
\ref{th:nonyu}, with its proof  in Appendix.

\begin{Th}\label{th:nonyu}
  For any choice of $\pi^\ast(y,\u)$, under Conditions (B1)-(B5), $\wh\bb$ satisfies
  \bse
  \sqrt{N}(\wh\bb - \bb) \to N\{\0, \A^{-1}\B(\A^{-1})\trans\},
  \ese
  in distribution when $N\to\infty$,
  where
  \bse
  \A&=&E\left[\frac{d\bS\eff^\ast\{\X_i,R_i,R_iY_i,\bb,\b^\ast(\cdot,\bb)\}}{d\bb\trans}\right],\\
\B&=&\var[\bS\eff^\ast\{\X_i,R_i,R_iY_i,\bb,\b^\ast(\cdot,\bb)\}-\h(\X_i)],\\
\h(\x_i)&=&
\int\left[\{\pi(y,\u)-1\}\{
\v_i(y,\u)-{\cal A}_i(\b^\ast)(y,\u)\}+{\cal A}^{-1}\{\v_i- {\cal A}_i(\b^\ast)\}(y,\u)u_1(y,\u)
\right]d\mu(y,\u).
\ese
\end{Th}
Here
\bse
&&\frac{d\bS\eff^\ast\{\X_i,R_i,R_iY_i,\bb,\b^\ast(\cdot,\bb)\}}{d\bb\trans}\\
&\equiv&
\frac{\partial\bS\eff^\ast\{\X_i,R_i,R_iY_i,\bb,\b^\ast(\cdot,\bb)\}}{\partial\bb\trans}
+\frac{\partial\bS\eff^\ast\{\X_i,R_i,R_iY_i,\bb,\b^\ast(\cdot,\bb)\}}{\partial{\b^\ast}\trans}\frac{\partial
  \b^\ast(\cdot,\bb)}{\partial\bb\trans}.
\ese

\begin{Rem}\label{rem:empnon}
Theorem \ref{th:nonyu} has a similar $h(\x_i)$ term as in Theorem \ref{th:emp}.
Similar to Remark \ref{rem:emp}, this term can also be viewed as the
additional cost from replacing the integral in
(\ref{eq:intgeneral}) with its approximation in (\ref{eq:appintgeneral}).
\end{Rem}

\begin{Rem}\label{rem:disU}
So far in this Section, we have implicitly assumed that $\U$ is continuous.
When $\U$ contains discrete component, we only need to stratify the
data according to the different discrete values, then construct the
corresponding integral equations within each stratum according to
either (\ref{eq:appintspecial}) or
 (\ref{eq:appintgeneral}). Solving these integral equations will then
 provide $\wh\b^*(y,\u,\bb)$ and the remaining estimation procedures
 are completely identical to the last two steps in Algorithm 2.
Specifically, for discrete $\U$,
assume that $\U$ can be $\u_k^0, k=1, \dots, K$. Then, we replace
(\ref{eq:appintspecial}) with
\bse
&&\frac1{N_k}\sum_{i=1,\u_i=\u_k^0}^N
\left[
\frac{\f_\bb(y,\u_k^0,\z_i;\bb)}{\fyx(y,\u_k^0,\z_i;\bb)}
 -\b(y,\u_k^0)
+\frac{\int \f_\bb(t,\u_k^0,\z_i;\bb)\pi^\ast(t,\u_k^0)dt}{\int
  \fyx(t,\u_k^0,\z_i;\bb)\{1-\pi^\ast(t,\u_k^0)\}dt}
\right.\n\\
&&\left.
  -\frac{\int\b(t,\u_k^0)\fyx(t,\u_k^0,\z_i;\bb)\pi^\ast(t,\u_k^0)dt}{\int
  \fyx(t,\u_k^0,\z_i;\bb)\{1-\pi^\ast(t,\u_k^0)\}dt}\right]
\fyx(y,\u_k^0,\z_i;\bb)
=\0,
\ese
where $N_k=\sumI I(\u_i=\u_k^0)$, and solve it to obtain
$\wh\b^*(y,\u_k^0,\bb)$.
If $\U$ is a mix of discrete ($\U_d$) and continuous ($\U_c$) variables, say
$\U=(\U_d\trans, \U_c\trans)\trans$. Assume that
$\U_d$ can be $\u_{dk}^0, k=1, \dots, K$.
We then replace
(\ref{eq:appintgeneral}) with
\bse
\frac{1}{N_k}\sum_{i=1,\u_{di}=\u_{dk}^0}^N&&
\left[
\frac{\f_\bb(y,\u_{dk}^0,\u_{ci},\z_i;\bb)}{\fyx(y, \u_{dk}^0,\u_{ci},\z_i;\bb)}
 -\b(y, \u_{dk}^0,\u_{ci})\right.\\
&&\left.
+\frac{\int \f_\bb(t, \u_{dk}^0,\u_{ci},\z_i;\bb)\pi^\ast(t, \u_{dk}^0,\u_{ci})dt}{\int
  \fyx(t, \u_{dk}^0,\u_{ci},\z_i;\bb)\{1-\pi^\ast(t,
  \u_{dk}^0,\u_{ci})\}dt}
\right.\n\\
&&\left.
-\frac{\int\b(t, \u_{dk}^0,\u_{ci})\fyx(t, \u_{dk}^0,\u_{ci},\z_i;\bb)\pi^\ast(t, \u_{dk}^0,\u_{ci})dt}{\int
  \fyx(t, \u_{dk}^0,\u_{ci},\z_i;\bb)\{1-\pi^\ast(t, \u_{dk}^0,\u_{ci})\}dt}\right]\\
&&\times\fyx(y, \u_{dk}^0,\u_{ci},\z_i;\bb)K_h(\u_{ci}-\u_c)=\0,
\ese
where $N_k=\sumI I(\u_{di}=\u_{dk}^0)$, and solve it to obtain
$\wh\b^*(y,\u_{dk}^0,\u_c,\bb)$.
\end{Rem}

\section{Other Estimators}\label{sec:other}
In Sections \ref{sec:special} and
\ref{sec:general},
we proposed estimator $\wh\bb$ with minimum assumption regarding
estimating or modeling $\fx(\x)$
and $\fzu(\z,\u)$. If we are willing and able to
adopt further modeling and estimation procedures
to assess $\fx(\x)$ and $\fzu(\z,\u)$, different estimators for $\bb$
can be obtained. We illustrate two
alternative estimators.

Firstly, instead of approximating the expectations empirically,
we can use nonparametric kernel method in both
Sections \ref{sec:special} and \ref{sec:general}.
For example, in Section \ref{sec:special}
we can approximate $\fx(\cdot)$ via
$\wh f_\X(\x)=N^{-1}\sumI K_h(\x_i-\x)$, then
insert it into (\ref{eq:intspecial}) to form an approximate integral
equation. We denote the resulting
estimator $\wh\bb\non$. We summarize its property below, with its
proof  in Appendix. 

\begin{Th}\label{th:non}
  For any choice of $\pi^\ast(y)$, under Conditions (A1)-(A3), if
  $Nh^{2m}\to0$, then $\wh\bb\non$ satisfies
  \bse
  \sqrt{N}(\wh\bb\non - \bb) \to N\{\0, \A\non^{-1}\B\non(\A\non^{-1})\trans\},
  \ese
  in distribution when $N\to\infty$,
  where
  \bse
  \A\non&=&E\left[\frac{d\bS\eff^\ast\{\X_i,R_i,R_iY_i,\bb,\b^\ast(\cdot,\bb)\}}{d\bb\trans}\right],\\
\B\non&=&\var[\bS\eff^\ast\{\X_i,R_i,R_iY_i,\bb,\b^\ast(\cdot,\bb)\}-\h(\X_i)],\\
\h(\x_i)&=&
\int\left[\{\pi(y)-1\}\{
\v_i(y)-{\cal A}_i(\b^\ast)(y)\}+{\cal A}^{-1}\{\v_i- {\cal A}_i(\b^\ast)\}(y)u_1(y)
\right]d\mu(y).
\ese
\end{Th}
Here
\bse
&&\frac{d\bS\eff^\ast\{\X_i,R_i,R_iY_i,\bb,\b^\ast(\cdot,\bb)\}}{d\bb\trans}\\
&\equiv&
\frac{\partial\bS\eff^\ast\{\X_i,R_i,R_iY_i,\bb,\b^\ast(\cdot,\bb)\}}{\partial\bb\trans}
+\frac{\partial\bS\eff^\ast\{\X_i,R_i,R_iY_i,\bb,\b^\ast(\cdot,\bb)\}}{\partial{\b^\ast}\trans}\frac{\partial
  \b^\ast(\cdot,\bb)}{\partial\bb\trans}.
\ese

\begin{Rem}
Similar to Theorem \ref{th:non}, under the assumption
(\ref{eq:general}) in Section \ref{sec:general}, a pure nonparametric
kernel based estimator can also be derived. We omit the details to
avoid repetition.
\end{Rem}

\begin{Rem}
Similar to the discussion in Section \ref{sec:general}, in the above analysis of $\wh\bb\non$, we
have assumed that all components in $\X$ are continuous.
If $\X$ contains discrete components, say
$\X=(\X_c\trans, \X_d\trans)\trans$, where $\X_c$ consists of
continuous variables and $\X_d$ is the collection of discrete
variables, then we need to slightly
adjust the procedure. Specifically, let $\X_d$ have values $\x_{dk}^0,
k=1,\dots, K$.
We would stratify the data into $K$ strata.
 Within each stratum, we treat  $\X_c$ as the new $\X$
variable and write the kernel estimator $\wh f_{\X_c\mid \X_d =
  \x_{dk}^0}$ as $\wh f_{\X_c,k}$. The integral equation
(\ref{eq:intspecial}) is then approximated by
\bse
&&\sum_{k=1}^K \wh p_k\int \left\{\f_\bb(y,\x_c, \x_{dk};\bb)\wh
  f_{\X_c,k}(\x_c)\right.\\
&&\left.+\frac{\int\f_\bb(t,\x_c, \x_{dk};\bb)\pi^\ast(t)dt}{1-\int
  \fyx(t,\x_c, \x_{dk};\bb)\pi^\ast(t)dt}\fyx(y,\x_c, \x_{dk};\bb)\wh f_{\X_c,k}(\x_c)\right\}d\mu(\x_c)\\
&=&\sum_{k=1}^K \wh p_k
\int\left\{\b(y)\fyx(y,\x_c, \x_{dk};\bb)\wh f_{\X_c,k}(\x_c)\right.\\
&&\left.+\frac{\int\b(t)\fyx(t,\x_c, \x_{dk};\bb)\pi^\ast(t)dt}{1-\int
  \fyx(t,\x_c, \x_{dk};\bb)\pi^\ast(t)dt}\fyx(y,\x_c, \x_{dk},\bb)\wh f_{\X_c,k}(\x_c)\right\}d\mu(\x_c),\n
\ese
where $\wh p_k$ is the empirical frequency of observations in the
$k$th stratum. After solving the integral equation, we still proceed
to the same estimating equation in Algorithm 1.
\end{Rem}

Secondly, we consider  parametric estimation of $\fx(\x)$ and
$\fzu(\z,\u)$, i.e.
$\fx(\x;\wh\ba)$ in Section \ref{sec:special}  and
$\fzu(\z,\u;\wh\ba)$ in Section \ref{sec:general}.
This scenario can arise in the
situation when one is confident to correctly specify a
parametric model, using all fully observed data.
For convenience, we assume $N^{1/2}(\wh\ba-\ba)=N^{-1/2}\sumI\bphi(\x_i;\ba)+o_p(1)$,
which is the typical expansion for most full data parametric estimators.
For example, when maximum likelihood estimator (MLE) is used,
$\bphi(\x_i;\ba)=-[E\{{\partial^2 \log
  \fx(\x;\ba)}/{\partial\ba\partial\ba\trans}\}]^{-1} {\partial \log
  \fx(\x_i;\ba)}/{\partial\ba}$.
 We call the corresponding estimator $\wh\bb\para$.
For $\wh\bb\para$ we have the following asymptotic result and its
proof is in Appendix. 

\begin{Th}\label{th:par}
For both the special assumption (\ref{eq:special}) with  an arbitrary
choice of $\pi^\ast(y)$,
and the general
assumption (\ref{eq:general}) with an arbitrary choice of
$\pi^\ast(y,\u)$, the corresponding estimator $\wh\bb\para$
satisfies
 \bse
  \sqrt{N}(\wh\bb\para - \bb) \to N\{\0, \A\para^{-1}\B\para(\A\para^{-1})\trans\}
  \ese
in distribution when $N\to\infty$,
  where
  \bse
\A\para&=&E\left[\frac{d\bS\eff^\ast\{\X_i,R_i,R_iY_i,\bb,\b^*(\cdot,\bb,\ba)\}}{d\bb\trans}\right].
\ese
Under the special assumption (\ref{eq:special}),
\bse
\B\para&=&\var\left(\bS\eff^\ast\{\X_i,R_i,R_iY_i,\bb,\b^*(\cdot,\bb,\ba)\}\right.\\
&&\left.+
 E\left[
\bS\eff^\ast\{\X_i,R_i,R_iY_i,\bb,\b^*(\cdot,\bb,\ba)\}\frac{\partial\log f_\X(\X_i;\ba)}{\partial\ba\trans}\right]
\bphi(\X_i;\ba)
\right).
\ese
Under the general assumption (\ref{eq:general}),
$\B\para$ has the same form
but with ${\partial\log f_\X(\X_i;\ba)}/{\partial\ba\trans}$ replaced
by
${\partial\log \fzu(\Z_i,\U_i;\ba)}/{\partial\ba\trans}$.
\end{Th}

\begin{Rem}\label{rem:par}
In practice, a potential obstacle to using the parametric model and the result
of Theorem \ref{th:par} is the possible model
misspecification. Theorem \ref{th:par} shows that, when
this parametric model is indeed correct, the variance of the estimator
contains an additional term, which resembles $h(\x_i)$ in Theorems \ref{th:emp}
and \ref{th:nonyu}.
Furthermore, if the working model $\pi^*$ happens to be
  correctly specified, this term is zero.
\end{Rem}


\section{Simulation Studies}
\label{sec:simu}

We conduct simulation studies to evaluate the finite sample
performance of our proposed estimator $\wh\bb$ in Theorems \ref{th:emp} and \ref{th:nonyu}.
We mainly compare with the alternative estimators $\wh\bb\non$ and $\wh\bb\para$ presented in Section \ref{sec:other}.
To evaluate the performance against the theoretical optimal limit,
we  also implement the oracle estimator $\wh\bb\ora$,
obtained when the true $\fx(\x)$ in (\ref{eq:intspecial}), or $\fzu(\z,\u)$ in
(\ref{eq:intgeneral}), is used.
We first present the results under the special assumption (\ref{eq:special}),
then the results under the general assumption (\ref{eq:general}).

\subsection{Scenarios under Special Assumption (\ref{eq:special})}
\label{sec:simu1}

We experiment two situations under the special assumption (\ref{eq:special}).
In each situation, we consider eight different methods,
  where the working mechanism $\pi^*(y)$ is correct or misspecified,
  in combination with $\wh f_\X(\x)$ being obtained by one of the four
  approaches: its truth, the proposal in Section \ref{sec:special},
  and the two alternatives in Section \ref{sec:other}.


For the first situation, we generate $X$ from a univariate normal
distribution with mean 0.5 and variance $\sigma^2=0.25$. The response
$Y$ is generated from the model
$Y = \beta_0 + \beta_1 X + \epsilon$,
where the parameter of interest $\bb=(\beta_0, \beta_1)\trans=(0.25, -0.5)\trans$,
and $\epsilon$ follows the standard normal distribution. The true model of
the missingness mechanism is
\bse
\pr(R=1\mid y)=\pi(y)=\frac{\exp(1+y)}{1+\exp(1+y)}.
\ese
This leads to about 1/3 subjects with missing response. The
misspecified working mechanism model is
\bse
\pi^\ast(y)=\frac{\exp(1-y)}{1+\exp(1-y)}.
\ese

In terms of numerical implementation,
we use the Gauss-Hermite quadrature with $15$  points to
approximate the integrations. We
adopt the Epanechnikov kernel function
$K(u)=\frac34 (1-u^2)I_{\{|u|\leq 1\}}$ with $m=2$
in the nonparametric density estimation. We choose the bandwidth $C
N^{-1/3}$ with $C=1.5$ in our simulations.
We find that the results are
robust in the situations where $C$ ranges from 1 to 2.

We consider the total sample size $N=500$ and the results summarized
in Table~\ref{tb:part1} are based on 1,000 simulation replicates. For
each estimator, we compute its sample bias (\texttt{bias}), sample standard
derivation (\texttt{std}), estimated standard derivation using the asymptotic
distribution (\texttt{$\wh{\mbox{std}}$}) and the coverage probability
(\texttt{cvg}) at the nominal level 95\%.

\begin{center}
[Table~\ref{tb:part1} approximately here]
\end{center}

In the second situation, we generate $\X$ from a 3-dimensional
multivariate normal distribution with mean zero and covariance matrix
$\bSig=(0.5^{|i-j|})_{1\leq i,j\leq 3}$, and generate $Y$ from the
linear model
$Y=\beta_0 + \beta_1 X_1 + \beta_2 X_2 + \beta_3 X_3 + \epsilon$. Here
 $\bb=(\beta_0, \beta_1, \beta_2, \beta_3)\trans=(0, 0.1,
-0.2, -0.3)\trans$ and $\epsilon$ has the standard normal distribution.
We adopt the same missingness mechanism model as in the
univariate $X$ case and it also leads to  around 1/3 missingness.
The same misspecified working mechanism model and kernel function
are  used in estimation.
In implementing the
multivariate Gauss-Hermite quadrature \citep{jackel2005note} to
approximate the integrations, we adopt 6 quadrature points in each dimension
which generates $6^3=216$ points in total.
We set the bandwidth to $2N^{-2/7}$, and find the results
robust if the constant $2$ varies between 1.5 to 2.5.
We consider sample size $N=1,000$ and the
results summarized in Table \ref{tb:part2} are also based on 1,000
simulation replicates.

\begin{center}
[Table~\ref{tb:part2} approximately here]
\end{center}

We reach the following conclusions from summarizing the results in
Tables \ref{tb:part1} and \ref{tb:part2}.
First, in each scenario and for all the estimators we considered,
the biases are very close to zero, the sample standard deviation
and the estimated standard deviation are rather close to each other,
and the sample coverage rates of the estimated 95\% confidence intervals
are close to the nominal level 95\%.
Hence, regardless of how $\fx(\x)$ is estimated and whether the
working mechanism model $\pi^\ast(y)$
is specified correctly or not, our methodology always produces
consistent estimator and the inference results based on the asymptotic
results are sufficiently precise.
Second, in each of the scenarios considered, although the estimator
with a misspecified mechanism $\pi^\ast(y)$ is less efficient than its
counterpart with
the true $\pi(y)$, the inflation of the standard deviation is not
large. This indicates a certain robustness of our method to the
working missingness mechanism model in terms of estimation efficiency,
in addition to the established estimation consistency. This seems to
be an added advantage of our
estimator because
the true form of $\pi(y)$ is
difficult to obtain in practice. Our observation here helps
to alleviate the burden of extensive efforts to identify a proper
missingness mechanism description $\pi(y)$ in order to reach sufficiently small
estimation variability.
Third, when the true $\pi(y)$ is used, all estimators have similar
  numerical performance, especially in the $p=3$ case. Similar
phenomenon is also observed when $\pi(y)$ is misspecified.
Therefore, considering the possible model
misspecification of $\fx(\x)$ in $\wh\bb\para$ and the potential difficulty
of nonparametric estimation in implementing $\wh\bb\non$, we highly
 recommend the
use of $\wh\bb$ in practice.

\subsection{Scenarios under General Assumption (\ref{eq:general})}
\label{sec:simu2}

Under the general assumption (\ref{eq:general}), we also perform two different
simulation studies to examine the finite sample performance of our proposed
estimators.

In the first study, both $\U$ and $\Z$ are continuous
variables so the theory established in Theorem \ref{th:nonyu} applies.
We consider treating the conditional expectation related to the unknown quantity $\fx(\u,\z)$
via nonparametric regression, parametric modeling or adopting the truth,
in combination with the mechanism model being correct or
misspecified. Thus, we
implement six different estimators.

The data generation process is as
follows. We first generate $\X$ from a bivariate normal distribution
with mean zero and covariance matrix $\bSig =
(0.5^{|i-j|})_{1\leq i,j\leq 2}$. Then we generate the outcome $Y$
from
\bse
\logit\{\pr(Y=1\mid u,z)\} = \beta_0 + \beta_1 u + \beta_2 z
\ese
with the parameter of interest $\bb=(\beta_0, \beta_1, \beta_2)\trans
= (0, 0.3, -0.3)\trans$. The missing data indicator $R$ is generated
from
\bse
\pr(R=1\mid y,u)=\pi(y,u)=\frac{\exp(1+y+u)}{1+\exp(1+y+u)},
\ese
which yields approximately 20\% missingness in $Y$. We adopt the
misspecified working mechanism model as
\bse
\pi^*(y,u) = \frac{\exp(1-y-u)}{1+\exp(1-y-u)}.
\ese
With the total sample size $N=1,000$, we implement the estimator
$\wh\bb$ following Algorithm 2 in Section \ref{sec:general} and
the estimators $\wh\bb\ora$ and $\wh\bb\para$ following the discussion
in Section \ref{sec:other}.
We adopt the Epanechnikov kernel
in (\ref{eq:appintgeneral}).
Similar to Section \ref{sec:simu1}, we use the Gauss-Hermite
quadrature with
$15$ bases to
approximate the integrals.
The bandwidth is chosen as $C
N^{-1/3}$ with $C=2$. Results
based on 1,000 simulation are summarized in Table~\ref{tb:part3c}.

\begin{center}
[Table~\ref{tb:part3c} approximately here]
\end{center}

The second simulation study is designed to mimic the
real data example presented in Section \ref{sec:data}.
We first generate binary covariate $U$ from a
bernoulli distribution with $\pr(U=1)=0.5$.
Then we generate $Z$ following
\bse
\logit\{\pr(Z=1\mid u)\}=-1.5 + 0.2 u.
\ese
The outcome variable $Y$ is generated from
\bse
\logit\{\pr(Y=1\mid u,z)\}=\beta_0 + \beta_1 u + \beta_2 z
\ese
with $\bb=(\beta_0, \beta_1, \beta_2)\trans=(-0.5, 0.2, 0.7)\trans$.
We then generate the missing data indicator $R$ following
\bse
\pr(R=1\mid y,u)=\pi(y,u)=\frac{\exp(1 -2 y + 0.3 u)}{1+\exp(1 -2 y + 0.3 u)}.
\ese
We use the working model
\bse
\pi^\ast(y,u)=\frac{\exp(1 +2 y + 0.3 u)}{1+\exp(1 +2 y + 0.3 u)}
\ese
as the misspecified mechanism model.
We also implement  the six different estimators,
  respectively  $\wh\bb$, $\wh\bb\ora$ and $\wh\bb\para$ in
  combination with a correct or misspecified mechanism model.
Results based on sample size $N=2,000$ and 1,000
simulation replications are privided in Table \ref{tb:part3}.

\begin{center}
[Table~\ref{tb:part3} approximately here]
\end{center}

The conclusions from summarizing Tables~\ref{tb:part3c} and \ref{tb:part3} are also very
clear. First, similar to Section \ref{sec:simu1}, regardless of how
$\fx(\u,\z)$ is estimated and whether $\pi(y,\u)$ is specified
correctly or not, our methods always produce consistent estimators
and the inference results based on the asymptotic results are
sufficiently precise. Second, the estimator with an incorrect
$\pi(y,\u)$ is less efficient than its counterpart with the
correct $\pi(y,\u)$ model, while the efficiency loss is large.
Third, when the true $\pi(y,\u)$ model is
used, the estimators $\wh\bb$ and $\wh\bb\para$
perform similarly and they are both slightly less efficient than
$\wh\bb\ora$. The same phenomenon is observed when the misspecified
$\pi(u,\u)$ model is used.
All of these phenomena  match
our theory investigated in Sections \ref{sec:general} and
\ref{sec:other} very closely. In conclusion, in view of the
risks involved in a parametric model $\fzu(\z,\u)$ and the
infeasibility of the oracle estimator, we highly recommend using the
proposed estimator $\wh\bb$.

\section{Real Data Analysis}
\label{sec:data}

\cite{ibrahim2001using} analyzed a data set of mental health of
children in Connecticut \citep{zahner1992children, zahner1993rural,
  zahner1997factors}, where the binary outcome of interest is the
teacher's report of the psychopathology of the child (a score of 1
indicates borderline or clinical psychopathology, and a score of 0
indicates normal). The two covariate variables of interest are
\verb"father", the parental status of the household (0 indicates
father figure present, and 1 no father figure present), and
\verb"health", the physical health of the child (0 means no health
problems, and 1 means fair or poor health, a chronic condition or a
limitation in activity). In this study, a child's possibly unobserved
psychopathology status may be related to missingness because a teacher
is more likely to fill out the psychopathology status when the teacher
feels that the child is normal or not normal. Hence it is highly
suspected that the missingness mechanism is nonignorable. There are
2,486 subjects in this data set and 1,061 of them have missing outcome
values. The data set is available in \cite{ibrahim2001using}.

\cite{ibrahim2001using} firstly implemented a parametric EM algorithm
(the method \verb"parEM") where the mechanism is a logistic regression
model. Interestingly they found \verb"father" to be unrelated to
missingness, therefore they dropped out the covariate \verb"father" in
the mechanism model in their downstream analysis. Then what
\cite{ibrahim2001using} proposed to do is to replace the outcome
variable (teacher's report) in the mechanism model as the parents'
report of the psychopathology of the child, a completely observed
auxiliary variable, so that an ignorable missingness mechanism model,
instead of nonignorable, can be explored (the method \verb"ILH"). \cite{ibrahim2001using}
examined that, if the auxiliary variable is highly correlated to the
outcome variable, the method \verb"ILH" could reduce the estimation
bias, compared to the naive method using only completely observed
subjects (the method \verb"CC").

Since \verb"father" was found to be unrelated to missingness, it
could serve as the shadow variable in our context, therefore we apply
our proposed methodology in this study.
We implement the proposed estimator $\wh\bb$,
and the estimator $\wh\bb\para$ where $\fzu(\u,\z)$ is
modeled as
\bse
\logit\{\pr(\texttt{father=1}\mid \texttt{health})\} = -1.421 + 0.159\ \texttt{health}.
\ese
The posited missingness mechanism
model $\pi^*(y,\texttt{health})$ used in both $\wh\bb$ and $\wh\bb\para$ is
\bse
\pr(R=1\mid y, \texttt{health}) = \frac{\exp(1.013 - 2.139 y + 0.303\ \texttt{health})}{1+ \exp(1.013 - 2.139 y + 0.303\ \texttt{health})},
\ese
the same as
in the method \verb"parEM", reported in \cite{ibrahim2001using}.
 For each of the parameter coefficients, we report the estimate,
 standard error, its corresponding z-statistic and p-value for the
 five methods in Table~\ref{tb:realdata}.

\begin{center}
[Table~\ref{tb:realdata} approximately here]
\end{center}

Interestingly all methods produce roughly the same coefficient estimate for the
shadow variable \verb"father", although the
proposed method $\wh\bb$ gives the smallest standard error hence
the most efficient.
The primary differences among the five methods occur in the
coefficients of \verb"intercept" and \verb"health".
The method \verb"CC" which only uses completely observed subjects, the
method \verb"parEM" which is confined to a purely parametric model
specification, and the method \verb"ILH" which uses some auxiliary
variable and approximates the nonignorable mechanism by an ignorable
one, are all highly suspected to result in estimation biases.
The estimator $\wh\bb\para$, where the parametric
$\fx(\u,\z)$ model could be misspecified,
provides very similar estimates as the proposed estimator
$\wh\bb$.
However, $\wh\bb\para$ has a relatively large
standard error, as similarly observed in method \verb"parEM".
In contrast, the proposed estimator $\wh\bb$ is not prone to
any possible $\fx(\u,\z)$ model misspecification and appears much more
efficient than the estimator $\wh\bb\para$ in this application.

\section{Discussion}
\label{sec:disc}

In this paper, motivated by the difficulty of correctly specifying and
directly estimating the nonignorable missingness mechanism, we propose
a class of estimators which only need a working mechanism
 model hence avoids its correct specification and
estimation. Our procedure will always guarantee an asymptotically
consistent estimate for the parameter of interest.

In practice,
a choice of the working model for the missingness mechanism closer to its truth would be beneficial. To propose such a good model, one  can first adopt a pure
parametric likelihood estimator and use  EM algorithm to identify a
plausible nonignorable missingness mechanism model. This mechanism
model can then be used
as the working model $\pi^*(y,\u)$ in our procedure.

To achieve identifiability,
a major assumption in our estimation procedure is the existence of
the shadow variable $\Z$.
From the example we show in this paper and some other similar
situations, the existence of such a variable is clinically reasonable
and practically useful. How to statistically validate a shadow
variable is also of interest and it warrants further research.

\section*{Acknowledgement}

\begin{table}
\centering
\caption{Under assumption (\ref{eq:special}), univariate $X$.
Sample bias (bias), sample standard derivation (std), estimated
standard derivation  ($\wh{\mbox{std}}$), and coverage probability
(cvg) of 95\% confidence intervals of oracle estimator $\wh\bb\ora$, the mainly proposed estimator $\wh\bb$ studied in Theorem \ref{th:emp}, the estimator $\wh\bb\non$ studied in Theorem \ref{th:non}, and the estimator $\wh\bb\para$ studied in Theorem \ref{th:par}.}
\begin{tabular}{lllrrr}
\hline
\hline
Method & $\fx(\x)$ & $\pi(y)$ & Measure & $\beta_0$ & $\beta_1$\\
\hline
\hline
\multirow{8}{*}{$\wh\bb\ora$} & \multirow{8}{*}{True} &\multirow{4}{*}{Correct}
& bias & -0.0186 & -0.0040 \\
& & & std & 0.2033 & 0.1217\\
& & & $\wh{\mbox{std}}$ & 0.1943 & 0.1240\\
& & & cvg & 0.9530 & 0.9500\\
\cline{3-6}
&& \multirow{4}{*}{Incorrect}
& bias & -0.0198 & -0.0049 \\
& & & std & 0.2089 & 0.1281\\
& & & $\wh{\mbox{std}}$ & 0.2079 & 0.1274\\
& & & cvg & 0.9520 & 0.9460\\
\hline
\hline
\multirow{8}{*}{$\wh\bb$} & \multirow{8}{*}{Empirical} & \multirow{4}{*}{Correct}
& bias & -0.0156 & -0.0040\\
& & & std & 0.2088 & 0.1271\\
& & & $\wh{\mbox{std}}$ & 0.2121 & 0.1289\\
& & & cvg & 0.9530 & 0.9510\\
\cline{3-6}
& & \multirow{4}{*}{Incorrect}
& bias & -0.0185 & -0.0052 \\
& & & std & 0.2210 & 0.1362\\
& & & $\wh{\mbox{std}}$ & 0.2176 & 0.1408\\
& & & cvg & 0.9580 & 0.9560\\
\hline
\hline
\multirow{8}{*}{$\wh\bb\non$} & \multirow{8}{*}{Nonparametric} & \multirow{4}{*}{Correct}
& bias & -0.0180 & -0.0044\\
& & & std & 0.2067 & 0.1257\\
& & & $\wh{\mbox{std}}$ & 0.2109 & 0.1308\\
& & & cvg & 0.9530 & 0.9520\\
\cline{3-6}
& & \multirow{4}{*}{Incorrect}
& bias & -0.0136 & -0.0042 \\
& & & std & 0.2248 & 0.1433\\
& & & $\wh{\mbox{std}}$ & 0.2201 & 0.1439\\
& & & cvg & 0.9580 & 0.9680\\
\hline
\hline
\multirow{8}{*}{$\wh\bb\para$} & \multirow{8}{*}{Parametric} & \multirow{4}{*}{Correct}
& bias & -0.0177  & -0.0045 \\
& & & std & 0.1968 & 0.1253 \\
& & & $\wh{\mbox{std}}$ & 0.2015 & 0.1226\\
& & & cvg & 0.9510 & 0.9450\\
\cline{3-6}
& & \multirow{4}{*}{Incorrect}
& bias & -0.0140 & -0.0057 \\
& & & std & 0.2365 & 0.1481\\
& & & $\wh{\mbox{std}}$ & 0.2355 & 0.1357\\
& & & cvg & 0.9560 & 0.9370\\
\hline
\hline
\end{tabular}\label{tb:part1}
\end{table}

\begin{table}
\centering
\caption{Under assumption (\ref{eq:special}),  3-dimensional $\X$.
Sample bias (bias), sample standard derivation (std), estimated
standard derivation ($\wh{\mbox{std}}$), and coverage probability
(cvg) of  95\% confidence intervals of oracle estimator $\wh\bb\ora$, the mainly proposed estimator $\wh\bb$ studied in Theorem \ref{th:emp}, the estimator $\wh\bb\non$ studied in Theorem \ref{th:non}, and the estimator $\wh\bb\para$ studied in Theorem \ref{th:par}.}
\begin{tabular}{lllrrrrr}
\hline
\hline
Method & $\fx(\x)$ & $\pi(y)$ & Measure & $\beta_0$ & $\beta_1$ & $\beta_2$ & $\beta_3$\\
\hline
\hline
\multirow{8}{*}{$\wh\bb\ora$} & \multirow{8}{*}{True} & \multirow{4}{*}{Correct}
& bias & -0.0053 & -0.0018 & -0.0038 & 0.0079\\
& & & std & 0.0892 & 0.0776 & 0.0750 & 0.0835\\
& & & $\wh{\mbox{std}}$ & 0.0905 & 0.0742 & 0.0768 & 0.0863\\
& & & cvg & 0.9541 & 0.9613 & 0.9469 & 0.9541\\
\cline{3-8}
& & \multirow{4}{*}{Incorrect}
& bias & 0.0249 & 0.0006 & 0.0070 & 0.0081\\
& & & std & 0.0982 & 0.0857 & 0.0935 & 0.0924\\
& & & $\wh{\mbox{std}}$ & 0.0846 & 0.1010 & 0.1035 & 0.0954\\
& & & cvg & 0.9558 & 0.9573 & 0.9624 & 0.9639\\
\hline
\hline
\multirow{8}{*}{$\wh\bb$} & \multirow{8}{*}{Empirical} & \multirow{4}{*}{Correct}
& bias & -0.0028 & -0.0024 & -0.0035 & 0.0047\\
& & & std & 0.0951 & 0.0872 & 0.0802 & 0.0854\\
& & & $\wh{\mbox{std}}$ & 0.0886 & 0.0781 & 0.0855 & 0.0877\\
& & & cvg & 0.9566 & 0.9586 & 0.9519 & 0.9494\\
\cline{3-8}
& & \multirow{4}{*}{Incorrect}
& bias & 0.0167 & 0.0076 & 0.0018 & 0.0024\\
& & & std & 0.1085 & 0.1011 & 0.1043 & 0.0946\\
& & & $\wh{\mbox{std}}$ & 0.0958 & 0.1038 & 0.1010 & 0.1017\\
& & & cvg & 0.9604 & 0.9624 & 0.9586 & 0.9543\\
\hline
\hline
\multirow{8}{*}{$\wh\bb\non$} & \multirow{8}{*}{Nonparametric} & \multirow{4}{*}{Correct}
& bias & -0.0060 & -0.0087 & -0.0028 & 0.0081\\
& & & std & 0.0909 & 0.0981 & 0.0814 & 0.1013\\
& & & $\wh{\mbox{std}}$ & 0.0823 & 0.0975 & 0.0855 & 0.1046\\
& & & cvg & 0.9652 & 0.9675 & 0.9530 & 0.9617\\
\cline{3-8}
&  & \multirow{4}{*}{Incorrect}
& bias & 0.0275 & 0.0043 & 0.0019 & 0.0070\\
& & & std & 0.1043 & 0.1075 & 0.0970 & 0.1033\\
& & & $\wh{\mbox{std}}$ & 0.1085 & 0.1132 & 0.1067 & 0.1052\\
& & & cvg & 0.9659 & 0.9692 & 0.9670 & 0.9626\\
\hline
\hline
\multirow{8}{*}{$\wh\bb\para$} & \multirow{8}{*}{Parametric} & \multirow{4}{*}{Correct}
& bias & -0.0018 & -0.0031 & -0.0016 & 0.0052\\
& & & std & 0.0833 & 0.0761 & 0.0779 & 0.0815\\
& & & $\wh{\mbox{std}}$ & 0.0796 & 0.0748 & 0.0779 & 0.0802\\
& & & cvg & 0.9426 & 0.9464 & 0.9559 & 0.9447\\
\cline{3-8}
& & \multirow{4}{*}{Incorrect}
& bias & 0.0249 & 0.0044 & 0.0021 & 0.0097\\
& & & std & 0.1049 & 0.1011 & 0.0981 & 0.0921\\
& & & $\wh{\mbox{std}}$ & 0.0949 & 0.1033 & 0.0909 & 0.0900\\
& & & cvg & 0.9650 & 0.9556 & 0.9540 & 0.9492\\
\hline
\hline
\end{tabular}\label{tb:part2}
\end{table}

\begin{table}
\centering
\caption{Under assumption (\ref{eq:general}), continuous $U$.
Sample bias (bias), sample standard derivation (std), estimated
standard derivation ($\wh{\mbox{std}}$), and coverage probability
(cvg) of  95\% confidence intervals of oracle estimator $\wh\bb\ora$, the mainly proposed estimator $\wh\bb$ studied in Theorem \ref{th:nonyu}, and the estimator $\wh\bb\para$ studied in Theorem \ref{th:par}.}
\begin{tabular}{lllrrrr}
\hline
\hline
Method & $\fzu(\z,\u)$ & $\pi(y,\u)$ & Measure & $\beta_0$ & $\beta_1$ & $\beta_2$\\
\hline
\hline
\multirow{8}{*}{$\wh\bb\ora$} & \multirow{8}{*}{True} & \multirow{4}{*}{Correct}
& bias & -0.0179 & 0.0167 & 0.0045\\
& & & std & 0.0813 & 0.1214 & 0.1175\\
& & & $\wh{\mbox{std}}$ & 0.0751 & 0.1288 & 0.1209\\
& & & cvg & 0.9520 & 0.9460 & 0.9580\\
\cline{3-7}
& & \multirow{4}{*}{Incorrect}
& bias & -0.0134 & 0.0089 & 0.0036\\
& & & std & 0.0901 & 0.1287 & 0.1195\\
& & & $\wh{\mbox{std}}$ & 0.0974 & 0.1335 & 0.1107\\
& & & cvg & 0.9490 & 0.9510 & 0.9530\\
\hline
\hline
\multirow{8}{*}{$\wh\bb$} & \multirow{8}{*}{Empirical} & \multirow{4}{*}{Correct}
& bias & -0.0158 & 0.0123 & 0.0043\\
& & & std & 0.0825 & 0.1231 & 0.1147\\
& & & $\wh{\mbox{std}}$ & 0.0800 & 0.1189 & 0.1161\\
& & & cvg & 0.9500 & 0.9450 & 0.9530\\
\cline{3-7}
& & \multirow{4}{*}{Incorrect}
& bias & -0.0167 & 0.0155 & 0.0055\\
& & & std & 0.0916 & 0.1262 & 0.1200\\
& & & $\wh{\mbox{std}}$ & 0.0979 & 0.1283 & 0.1240\\
& & & cvg & 0.9480 & 0.9510 & 0.9530\\
\hline
\hline
\multirow{8}{*}{$\wh\bb\para$} & \multirow{8}{*}{Parametric} & \multirow{4}{*}{Correct}
& bias & -0.0140 & 0.0150 & 0.0064\\
& & & std & 0.0910 & 0.1281 & 0.1164\\
& & & $\wh{\mbox{std}}$ & 0.0875 & 0.1310 & 0.1294\\
& & & cvg & 0.9600 & 0.9520 & 0.9560\\
\cline{3-7}
&& \multirow{4}{*}{Incorrect}
& bias & -0.0156 & 0.0077 & 0.0038\\
& & & std & 0.1096 & 0.1307 & 0.1340\\
& & & $\wh{\mbox{std}}$ & 0.0984 & 0.1284 & 0.1259\\
& & & cvg & 0.9610 & 0.9540 & 0.9570\\
\hline
\hline
\end{tabular}\label{tb:part3c}
\end{table}

\begin{table}
\centering
\caption{Under assumption (\ref{eq:general}), discrete $U$.
Sample bias (bias), sample standard derivation (std), estimated
standard derivation ($\wh{\mbox{std}}$), and coverage probability
(cvg) of  95\% confidence intervals of oracle estimator $\wh\bb\ora$, the mainly proposed estimator $\wh\bb$ studied in Theorem \ref{th:nonyu}, and the estimator $\wh\bb\para$ studied in Theorem \ref{th:par}.}
\begin{tabular}{lllrrrr}
\hline
\hline
Method & $\fzu(\z,\u)$ & $\pi(y,\u)$ & Measure & $\beta_0$ & $\beta_1$ & $\beta_2$\\
\hline
\hline
\multirow{8}{*}{$\wh\bb\ora$} & \multirow{8}{*}{True} & \multirow{4}{*}{Correct}
& bias & 0.0302 & -0.0020& 0.0108 \\
& & & std & 0.1188 & 0.1250 &0.0975\\
& & & $\wh{\mbox{std}}$ & 0.1258 & 0.1194 &0.1024\\
& & & cvg & 0.9460 & 0.9580 &0.9460\\
\cline{3-7}
& & \multirow{4}{*}{Incorrect}
& bias & 0.0459 &  0.0134  &  0.0121\\
& & & std & 0.1371  &  0.1387  &  0.1069\\
& & & $\wh{\mbox{std}}$ & 0.1429  &  0.1495  &  0.1091\\
& & & cvg & 0.9570 & 0.9670 & 0.9680\\
\hline
\hline
\multirow{8}{*}{$\wh\bb$} & \multirow{8}{*}{Empirical} & \multirow{4}{*}{Correct}
& bias & 0.0215 & -0.0022& 0.0091 \\
& & & std & 0.1194 & 0.1266 & 0.1037\\
& & & $\wh{\mbox{std}}$ & 0.1205 & 0.1231 & 0.0975\\
& & & cvg & 0.9510 & 0.9520 & 0.9610\\
\cline{3-7}
& & \multirow{4}{*}{Incorrect}
& bias & 0.0533 & 0.0185 & 0.0081\\
& & & std & 0.1327 & 0.1455 & 0.1003\\
& & & $\wh{\mbox{std}}$ & 0.1395 & 0.1438 & 0.0994\\
& & & cvg & 0.9490 & 0.9600 & 0.9590\\
\hline
\hline
\multirow{8}{*}{$\wh\bb\para$} & \multirow{8}{*}{Parametric} & \multirow{4}{*}{Correct}
& bias &  0.0247 & -0.0018 & 0.0097\\
& & & std & 0.1143 & 0.1288 & 0.1018\\
& & & $\wh{\mbox{std}}$ & 0.1209 & 0.1265 & 0.0905\\
& & & cvg & 0.9610 & 0.9540 & 0.9630\\
\cline{3-7}
& & \multirow{4}{*}{Incorrect}
& bias & 0.0386 & 0.0097 & 0.0089\\
& & & std & 0.1475 & 0.1487 & 0.1121\\
& & & $\wh{\mbox{std}}$ & 0.1382 & 0.1547 & 0.1094\\
& & & cvg & 0.9620 & 0.9580 & 0.9640\\
\hline
\hline
\end{tabular}\label{tb:part3}
\end{table}

\begin{table}
\centering
\caption{Comparison of the real data analysis results in the
  children's mental health study. CC is the method using only
  completely observed subjects. parEM is the method using the EM
  algorithm with a purely parametric model specification.
ILH is the method proposed in \cite{ibrahim2001using}.
$\wh\bb$ is the mainly proposed estimator studied in Theorem \ref{th:nonyu}. $\wh\bb\para$ is the estimator studied in Theorem \ref{th:par} but with possible $\fzu(\cdot)$ model misspecification.
}
\begin{tabular}{lrrrr}
\hline
\hline
Method & Measure & \verb"intercept" & \verb"health" & \verb"father"\\
\hline
\multirow{4}{*}{CC} & estimate & -1.7372 & 0.2465 & 0.5419 \\
& standard error & 0.1070 & 0.1380 & 0.1607\\
& z-statistic & -16.2358&1.7863&3.3724\\
& p-value & 0.0000&0.0740&0.0007\\
\hline
\multirow{4}{*}{parEM} & estimate & -0.6410 & 0.1150 & 0.5450\\
& standard error & 0.5170 & 0.1420& 0.1610\\
& z-statistic &-1.2398 &0.8099&3.3851\\
& p-value & 0.2150&0.4180&0.0007\\
\hline
\multirow{4}{*}{ILH} & estimate & -1.7030 & 0.1810 & 0.5120\\
& standard error & 0.1050 & 0.1370& 0.1580\\
& z-statistic &-16.2190 &1.3212&3.2405\\
& p-value &0.0000 &0.1864&0.0012\\
\hline
\multirow{4}{*}{$\wh\bb$} & estimate & -0.7182  &  0.2831  &  0.5406 \\
& standard error & 0.3357 & 0.0875 & 0.0776\\
& z-statistic & -2.1394 & 3.2354 & 6.9665\\
& p-value & 0.0324 & 0.0012 & 0.0000\\
\hline
\multirow{4}{*}{$\wh\bb\para$ } & estimate & -0.7208  &  0.2695  &  0.5433\\
& standard error & 0.5858 & 0.6238 & 0.1685\\
& z-statistic & -1.2305 & 0.4320 & 3.2243\\
& p-value & 0.2185 & 0.6657 & 0.0013\\
\hline
\hline
\end{tabular}\label{tb:realdata}
\end{table}

\setlength{\bibsep}{0.85pt}
{
\bibliographystyle{ims}
\bibliography{note}
}

\end{document}